\documentclass[pra,aps,twocolumn,superscriptaddress,floatfix,tightenlines]{
revtex4}

\usepackage{amsmath}
\usepackage{amsfonts}
\usepackage{amssymb}
\usepackage{verbatim}
\usepackage{graphicx}

\newcommand{\ket}[1]{\ensuremath{|#1\rangle}}
\newcommand{\bra}[1]{\ensuremath{\langle #1|}}

\newcommand{\op}[1]{\mbox{\boldmath $\hat{#1}$}}

\begin{document}

\title{Entanglement-free Heisenberg-limited phase estimation}

\author{B. L. Higgins}
\affiliation{Centre for Quantum Dynamics, Griffith University, Brisbane 4111,
Australia}
\author{D. W. Berry}
\affiliation{Centre for Quantum Computer Technology, Macquarie University,
Sydney 2109, Australia}
\author{S. D. Bartlett}
\affiliation{School of Physics, University of Sydney, Sydney 2006, Australia}
\author{H. M. Wiseman}
\affiliation{Centre for Quantum Dynamics, Griffith University, Brisbane 4111,
Australia}
\affiliation{Centre for Quantum Computer Technology, Griffith University,
Brisbane 4111, Australia}
\author{G. J. Pryde}
\affiliation{Centre for Quantum Dynamics, Griffith University, Brisbane 4111,
Australia}

\begin{abstract}
\end{abstract}

\maketitle

\textbf{Measurement underpins all quantitative science. A key example is the
measurement
of optical phase, used in length metrology and many other applications. Advances
in precision measurement have consistently led to important scientific
discoveries. At the fundamental level, measurement precision is limited by the
number $N$ of quantum resources (such as photons) that are used. Standard
measurement schemes, using each resource independently, lead to a phase
uncertainty that scales as $1/\sqrt{N}$---known as the standard quantum limit.
However, it has long been conjectured \cite{Caves1981, Yurke1986} that it
should be possible to achieve a precision limited only by the Heisenberg
uncertainty principle, dramatically improving the scaling to $1/N$
\cite{Giovannetti2004}. It is commonly thought that achieving this
improvement requires the use of exotic quantum entangled states, such as the
NOON state \cite{Bollinger1996, Lee2002}. These states are  extremely difficult
to generate. Measurement schemes with counted photons or ions have been
performed with $N\leq6$ \cite{Rarity1990, Fonseca1999, Edamatsu2002,
Walther2004, Mitchell2004, Eisenberg2005, Leibfried2005, Sun2006, Nagata2007,
Resch2007}, but few have surpassed the standard quantum limit \cite{Nagata2007,
Leibfried2005} and none have shown Heisenberg-limited scaling. Here we
demonstrate experimentally a Heisenberg-limited phase
estimation procedure. We replace entangled input states with multiple
applications of the phase shift on unentangled single-photon states. We
generalize Kitaev's phase estimation algorithm \cite{Kitaev1996} using adaptive
measurement theory \cite{Wiseman1995, Berry2000, Armen2002, Cook2007} to achieve
a standard deviation scaling at the Heisenberg limit. For the largest number of
resources used ($N=378$), we estimate an unknown phase with a variance more
than 10 dB below the standard quantum limit; achieving this variance
would require more than 4,000 resources using standard interferometry. Our
results represent a drastic reduction in the complexity of achieving
quantum-enhanced measurement precision.}

Phase estimation is a ubiquitous measurement primitive, used for precision
measurement of length, displacement, speed, optical properties, and much more.
Recent work in quantum interferometry has focused on $n$-photon NOON
states \cite{Rarity1990, Fonseca1999, Lee2002, Edamatsu2002, Walther2004,
Mitchell2004, Eisenberg2005, Mitchell2005, Leibfried2005}, $\left(\ket{n}\ket{0}
+ \ket{0}\ket{n}\right)/\sqrt 2$, expressed in terms of number states of the two
arms of the interferometer. With this state, an improved phase sensitivity
results from a decrease in the phase period from $2\pi$ to $2\pi/n$. We achieve
improved phase sensitivity more simply using an insight from quantum computing.
We apply Kitaev's phase estimation algorithm \cite{Kitaev1996,Nielsen2000} to
quantum interferometry, wherein the entangled input state is replaced by
multiple passes through the phase shift. The idea of using multi-pass protocols
to gain a quantum advantage was proposed for the problem of aligning spatial
reference frames \cite{Rudolph2003}, and further developed in relation to clock
synchronization \cite{deBurgh2005} and phase estimation \cite{Giovannetti2006,
vanDam2007}.

\begin{figure*}
\centering\includegraphics[width=12cm]{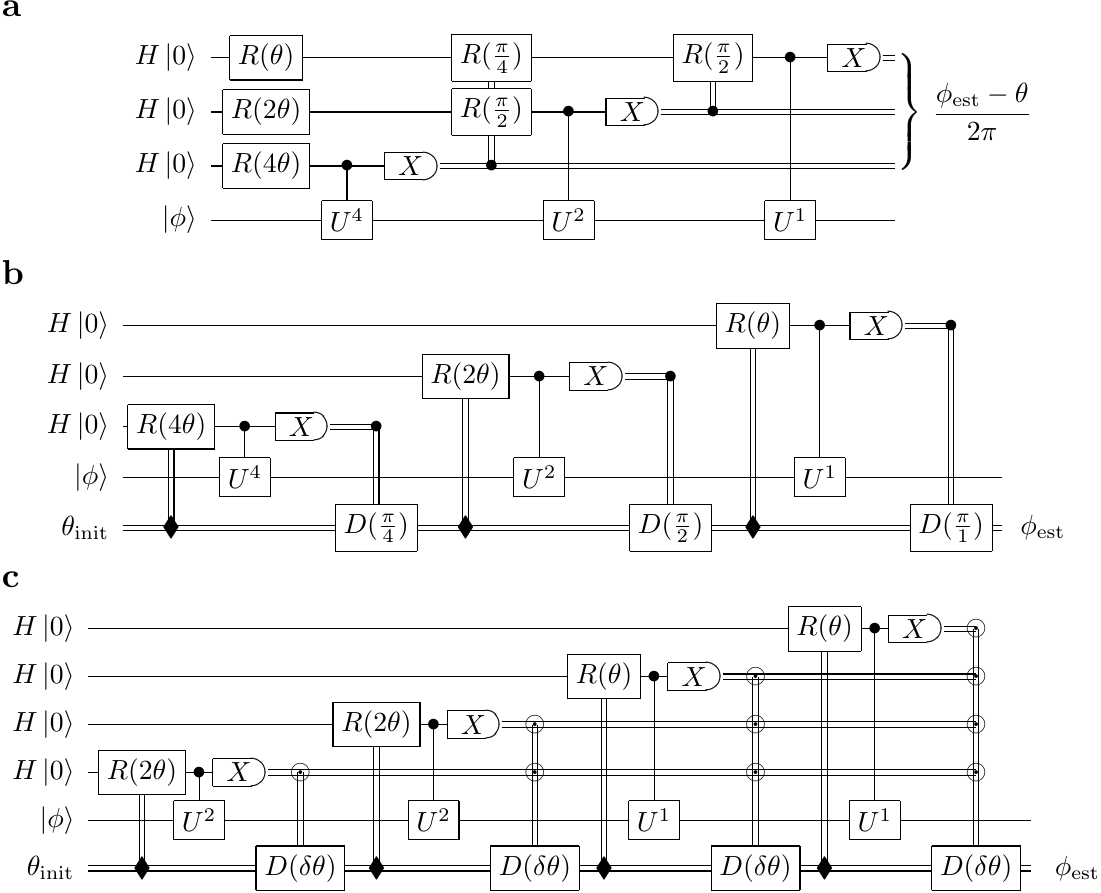}
\caption{\label{Fig:Theory} Quantum circuit diagrams of Kitaev's phase
estimation algorithm and our generalization. \textbf{a}, Kitaev's
algorithm \cite{Kitaev1996} with the inverse quantum Fourier transform
implemented with measurement and classical feedback \cite{Griffiths1996} and a
random initial phase estimate $\theta$. In general, $K+1$ qubits yield $K+1$
binary digits of precision; here $K=2$. \textbf{b}, As in \textbf{a}, but here
we implement $\theta$ (now called $\theta_\text{init}$) and the feedback
operations by coupling the qubits to a common element, the ``feedback phase''
$\theta$ (the lowest rail). \textbf{c}, Generalization of the circuit to include
$M\geq1$ qubits for each binary digit; here $K=1$ and $M=2$. For details on
circuit elements, see text.}
\end{figure*}

The conceptual circuit for Kitaev's phase estimation algorithm is shown in
Fig.~\ref{Fig:Theory}a. The algorithm  yields, with $K+1$ bits of precision, an
estimate $\phi_\text{est}$ of a classical phase parameter $\phi$, where
$e^{i\phi}$ is an eigenvalue of a unitary operator $U$. It requires us
to apply ${K+1}$ unitaries, $U^p$, with $p={2^K}, {2^{K-1}}, \ldots, 1$, each
controlled by a different qubit. Each qubit is prepared in the state $H\ket 0 =
\frac{1}{\sqrt{2}}\left(\ket 0 + \ket 1\right)$, and the control induces a phase
shift $e^{ip\phi}$ on the $\ket{1}$ component. The qubits are measured
sequentially in the $\op\sigma_x$ basis ($X$), and the results control
additional phase shifts, indicated by $R(\alpha) \equiv \exp(i\alpha
\ket{0}\bra{0})$, on subsequent qubits. This enables the inverse quantum Fourier
transform to be performed without entangling gates \cite{Griffiths1996}.
With a random phase $\theta$ on the qubits, as shown in Fig.~\ref{Fig:Theory}a,
the measurement results on the qubits are the binary digits of
$(\phi_\text{est}-\theta)/2\pi$; this ensures that the accuracy of the estimate
is independent of the value of $\phi$.

Alternatively, this independence could be obtained by using a second classical
``feedback'' phase $\theta$, as in Fig.~\ref{Fig:Theory}b, which also eliminates
the need for many of the gates in Fig.~\ref{Fig:Theory}a. This is a classical
real-valued parameter whose value is adjusted by $\pi/p$, indicated by the
symbol $D(\pi/p)$, controlled by the results of measurements. The value
of $\theta$ determines (as indicated by the diamond-shaped control symbol in
Fig.~\ref{Fig:Theory}b) phase-shifts $R(p\theta)$ on the qubits. Applying this
to interferometry, we can measure an unknown optical phase $\phi$ using
dual-rail photonic qubits \cite{Nielsen2000}. Here the operator $U$ induces a
relative phase shift $\phi$ each time the beam path (in one arm of the
interferometer) passes through the unknown optical phase $\phi$. The additional
phase shifts (determined by $\theta$) can be implemented using a single-pass
controllable phase in the other arm.

If a fixed probability of error in $\phi_\text{est}$ is allowed (that is, if the
uncertainty is quantified by a confidence interval), then the uncertainty of
Kitaev's phase estimation scales as $2^{-K}$ \cite{Nielsen2000}. Because
the number of control photons is $N_\text{phot}=K+1$, this scaling
implies an exponential decrease in the phase uncertainty with increasing
resources---apparently violating the Heisenberg uncertainty principle. The
correct analysis, however, is as follows. Although the cost of implementing
$U^{p}$ can be assumed to be essentially independent of $p$ in the context of
quantum computation, in interferometry it requires $p$ applications of the phase
shift, and should thus be counted as requiring $p$
resources \cite{Giovannetti2006}. Using this definition, the total number of
resources used is $N = 2^{K+1}-1$. Then for $N\gg1$, the uncertainty scales as
$1/N$, as in the Heisenberg limit. We note that this quantification of resources
in terms of the number of applications of the phase shift is the relevant one
for phase estimation of sensitive (for example, biological) samples, wherein the
goal is to pass as little light through the sample as is necessary.

On the other hand, if $\Delta\phi_\text{est}$ is taken to be the standard
deviation---the usual measure of uncertainty---then Kitaev's algorithm does
not scale as $1/N$. Rather, we have shown analytically that it asymptotes as
$\sqrt{2}/\sqrt N$, the same scaling as the standard quantum limit (SQL)---see
also Ref.~\cite{Mitchell2005}. The broad wings of the distribution of phase
estimates are not due to any deficiency in the estimation procedure---the
quantum Fourier transform is optimal---but rather are a consequence of the
sequence of phase shifts on the photons, $2^K \phi, 2^{K-1}\phi, \ldots, \phi$.

A key idea to address this problem is to employ $M$ copies of the control photon
at each phase shift \cite{Rudolph2003, deBurgh2005, Giovannetti2006}. For $M>1$
one cannot perform an exact quantum Fourier transform using only single-photon
operations. However, one can perform it approximately using the adaptive phase
estimation scheme of Ref.~\cite{Berry2000}, as shown in Fig.~\ref{Fig:Theory}c
for $M=2$. Here the feedback phase is adjusted by an amount $\delta\theta$,
controlled by all previous measurement results via a bayesian algorithm. This
general multi-bit conditioning is represented by the circled-dot symbol in
Fig~\ref{Fig:Theory}c. For the final adjustment, $\delta\theta = \phi_\text{est}
- \theta$, so the value of $\theta$ that is read out is equal to
$\phi_\text{est}$, the best estimate of the phase given all the data, as in
Fig.~\ref{Fig:Theory}b. Because the inverse quantum Fourier transform is not
performed exactly, and because the sequence of phase shifts is not exactly
equivalent to the optimal state of Ref.~\cite{Berry2000}, we do not expect
$\phi_\text{est}$ to have an uncertainty precisely at the Heisenberg limit
$\Delta\phi_\text{HL} = \tan\left[\pi/\left(N+2\right)\right] \approx \pi/N$ for
$N \gg 1$ \cite{Wiseman1997, Berry2000}.  Nevertheless, our algorithm
allows estimation with an uncertainty only a constant factor larger than this
ultimate limit, for $M\geq 4$. For instance, we find by numerical simulation
that for $M=6$, $\Delta\phi_\text{est} \approx 1.56\pi/N$ for $N \gg 1$.

\begin{figure}
\centering\includegraphics[width=\linewidth]{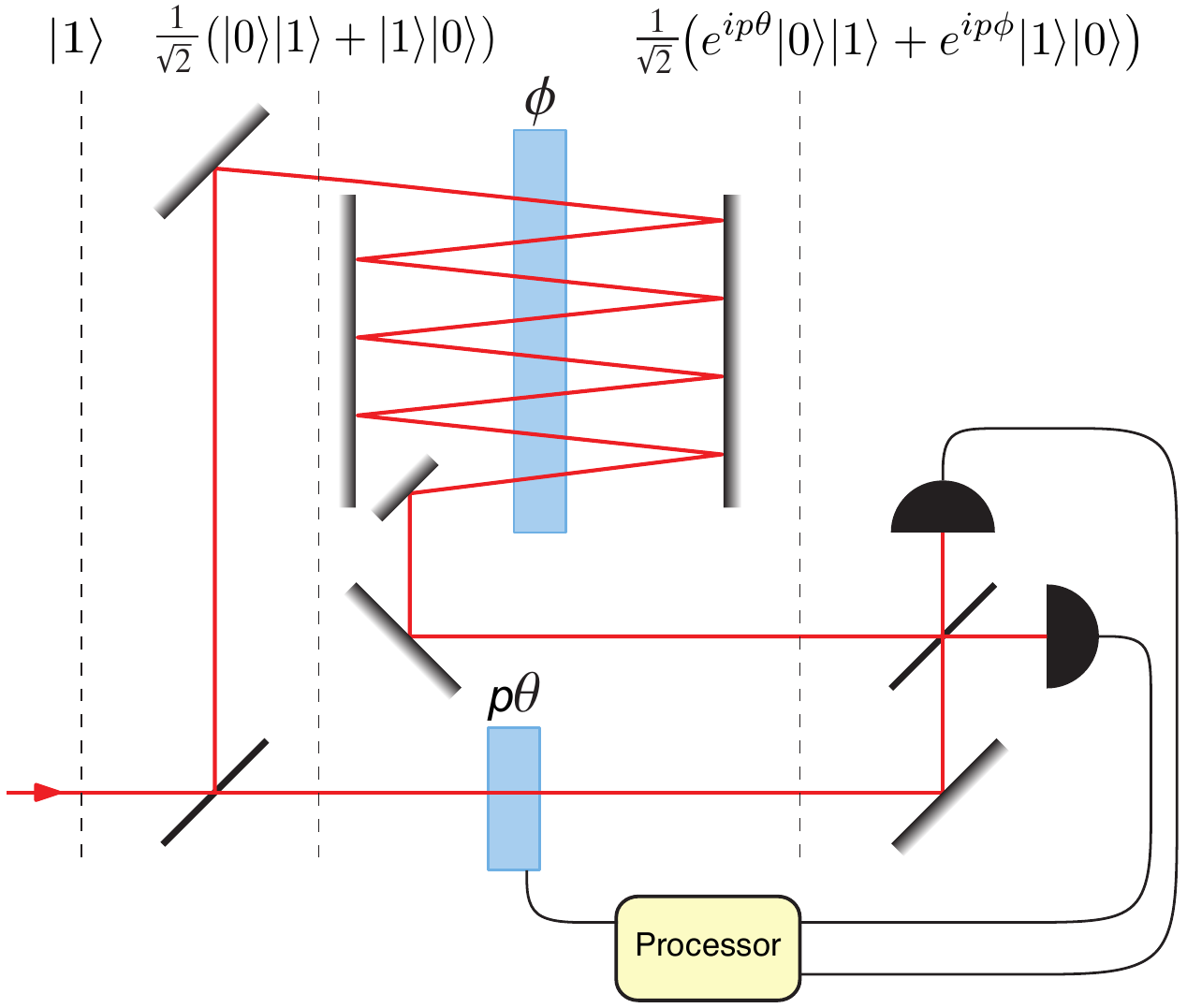}
\caption{\label{Fig:Concept} Conceptual diagram of the algorithm's
implementation as a Mach-Zehnder interferometer. This is equivalent to the
scheme in Fig.~\ref{Fig:Theory}c. Photon-number quantum states are shown at key
points. The first beam splitter implements the Hadamard operation on incident
photons. The large phase-shift element is configured to implement a
$p\phi$ phase shift on logical $\ket 1$ states, with $p$ adjustable ($p=8$
shown). The small phase-shift element implements the adjustable $p\theta$ phase
shift on logical $\ket 0$ states. The final beam splitter and single photon
detectors implement a $\op\sigma_{x}$ measurement, which determines, via the
processor, how to adjust $\theta$ before the next photon input.}
\end{figure}

A conceptual implementation of this generalization of Kitaev's algorithm is
shown in Fig.~\ref{Fig:Concept}. It works as follows: a photon is converted to
the state $\frac{1}{\sqrt{2}}\left({\ket 1}{\ket 0}+{\ket 0}{\ket 1}\right)$ by
the first beam splitter. After passing $p=2^K$ times through the phase shift
$\phi$, the state evolves to $\frac{1}{\sqrt{2}}\left(e^{ip\theta}{\ket 1}{\ket
0}+e^{ip\phi}{\ket 0}{\ket 1}\right)$. The photon is then detected after the
modes are recombined on the second beam splitter. The result is used to update
the probability distribution $P(\phi)$ which represents knowledge about $\phi$.

This process is repeated $M$ times, so that $M$ independent photons go through
$2^K$ passes in sequence. Quantifying a resource as a single pass of a
photon through the phase shift, each photon in this stage corresponds to $2^K$
resources. Following these $M$ photons, another $M$ photons undergo the same
process at $p=2^{K-1}$ passes, and so on for $p=2^{K-2},\ldots,2^0$. Thus a
total of $M\left(K+1\right)$ photons and $N=M\left(2^{K+1}-1\right)$
resources are used. The value of the feedback phase $\theta$ is random
for the first photon only. Thereafter it is chosen, based upon $P$ (that is,
upon all preceding results), to minimize the expected phase variance after the
next detection \cite{Berry2000}.

This bayesian control algorithm reduces to Kitaev's algorithm for $M=1$. We have
shown analytically that this algorithm gives a standard deviation of estimates
scaling as the SQL for $M = 2$ as well as $M=1$, but numerical simulations (for
$N$ up to $4\times10^6$ and $M$ up to 8) demonstrate a Heisenberg-limited
scaling for $M\geq4$.

We note that a single photon with $p$ passes through the unknown phase shift is
operationally equivalent to a NOON state with $n = p$ photons, and involves
exactly the same number of resources. A single NOON state such as this yields at
most one bit of information \cite{Bollinger1996}, and only about $\phi$ modulo
$2\pi/n$. It has been shown numerically \cite{Mitchell2005} that a sequence of
NOON states, with $n$ as well as $\theta$ chosen adaptively, achieves
Heisenberg-limited scaling, but only for $N > 100$. Our generalized algorithm,
which is simpler, can also be applied to NOON states, and directly achieves
Heisenberg-limited scaling. Even if high-$n$ NOON states could be produced,
however, they require high-$n$ photon-number-resolving detectors, and are
proportionately more sensitive to detector inefficiency than single photons with
multiple passes through the phase shift.

\begin{figure}
\centering\includegraphics[width=\linewidth]{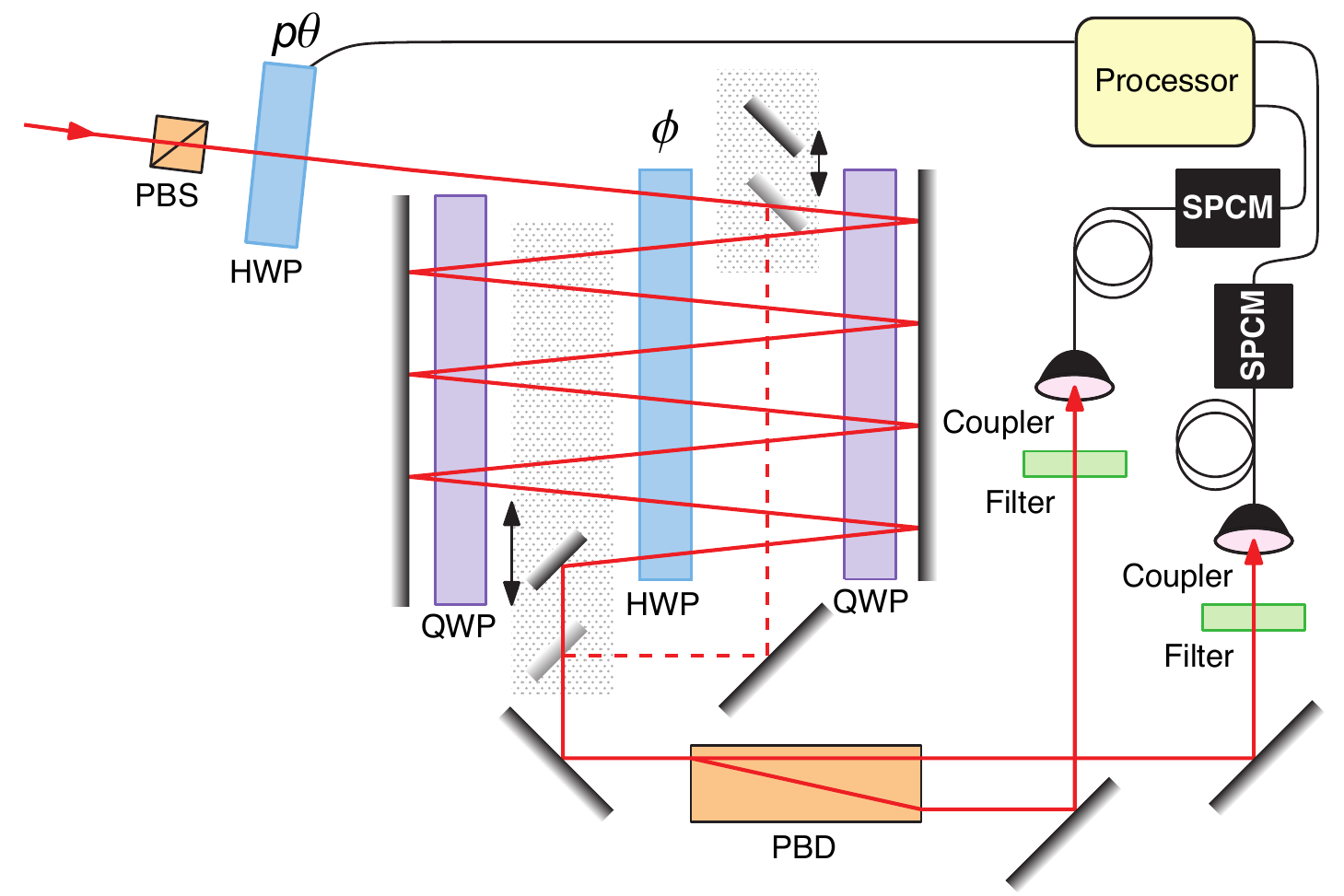}
\caption{\label{Fig:Experiment} Schematic of the experiment. Polarization modes
replace the arms of the interferometer in Fig.~\ref{Fig:Concept}, with phase
shifts implemented by half-wave plates (HWPs). A photon experiences phase shifts
between left- and right-circular polarizations by the feedback wave plate
($p\theta$) and the unknown phase wave plate ($\phi$). The photon is selected by
a mirror mounted on a motorized translation stage, discriminated in the
horizontal/vertical polarization basis by a polarizing beam displacer (PBD), and
passed through a 10-nm-bandwidth interference filter. It is then coupled into
a multimode fibre and detected by a single-photon counting module (SPCM),
completing the $\op\sigma_x$ measurement. PBS, polarizing beam splitter; QWP,
quarter-wave plate.}
\end{figure}

We demonstrate our single-photon algorithm using a common-spatial-mode
polarization interferometer, as shown in Fig.~\ref{Fig:Experiment}.
Common-spatial-mode interferometers are used in many metrology tasks involving
birefringent materials such as stress sensors, Faraday spectroscopy, and testing
optical components, but we stress that our algorithm applies equally well to any
interferometer. The two arms of the interferometer are the right-circular ($\ket
R$) and left-circular ($\ket L$) polarization modes. The unknown phase $\phi$ is
implemented as a birefringent half-wave plate. We test two versions of the
algorithm: $M=1$ (Kitaev's algorithm), and $M=6$ (chosen for its robustness).
For each, we vary the number of resources $N=M\left(2^{K+1}-1\right)$ by
choosing different values for the maximum number of passes, $2^K$, with
$K\in\{0,1,2,3,4,5\}$. We also measured the standard deviation for a
non-adaptive or ``standard'' estimation algorithm, using $N$ single passes
of the phase shift, with $N$ chosen equal to the number of resources used for
each of the $M=6$ data points. In this case $\theta$ was incremented
non-adaptively \cite{Hradil1996} by $\pi/N$ from one photon to the next, to
ensure a sensitivity independent of $\phi-\theta_\text{init}$.

\begin{figure}
\centering\includegraphics[width=\linewidth]{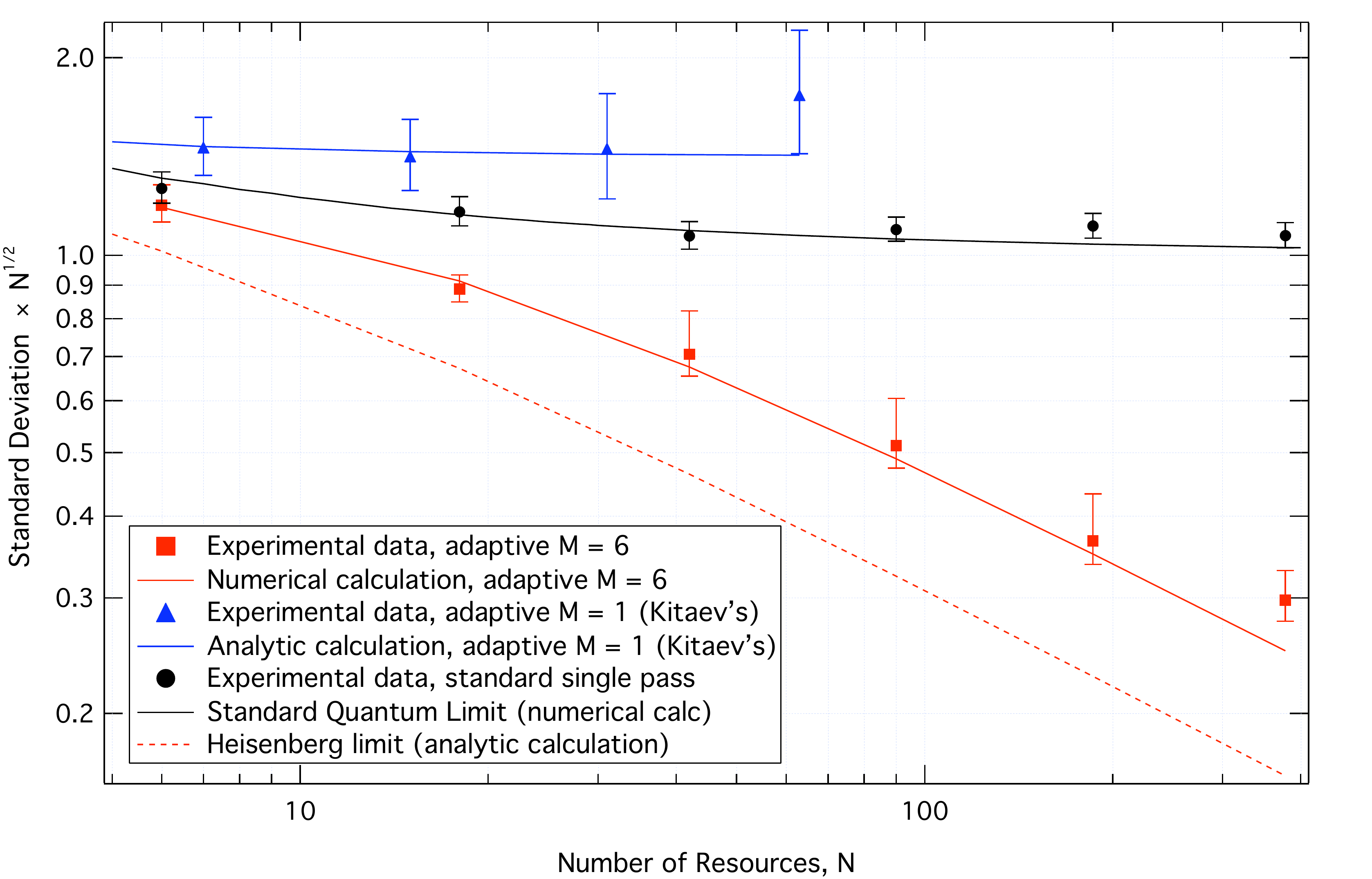}
\caption{\label{Fig:Results} Standard deviations of distributions of phase
estimates for varying numbers of resources $N$. We compare theoretical
predictions (lines) and measured values (points, each representing 1,000
estimates) for standard phase estimation and our implementation of Kitaev's
($M=1$) and generalized Kitaev's ($M=6$) algorithms. Error bars denote 95\%
confidence intervals. Our algorithm clearly has a lower standard deviation in
phase estimates than both the SQL and Kitaev's algorithm (which has SQL
scaling). For large $N$, the curve for the adaptive algorithm is parallel to the
Heisenberg limit, with a small overhead factor of about $1.56$.}
\end{figure}

The experimental results are shown in Fig.~\ref{Fig:Results}, together with
theoretical calculations. The error bars are 95\% confidence intervals
determined using a studentized bootstrap on a log scale \cite{Davison1997}. In
general, the distributions have a large positive kurtosis which emphasizes the
effect of outliers; our error calculation takes this into account to provide
accurate error bars. Theoretical predictions assume 100\% visibility.
Experimentally, visibilities for $p=1$ to $16$ were high (all above 98.1\% and
typically above 99.6\%). However the $p=32$ case had slightly lower visibility
(95.4\%), leading to higher than expected standard deviations for the $N=378$
case. This is primarily due to expansion of the beam, with consequent overlap
of beams, leading to a small probability of measuring the photon after only 30
passes.

The results of the non-adaptive phase estimation algorithm follow the SQL, as
expected. We note that the standard deviations of the $M=1$ (Kitaev's) case
also follow an SQL scaling. Most importantly, there is a clear Heisenberg
scaling, $\Delta\phi_\text{est}\propto1/N$, of our adaptive multi-pass algorithm
for $M=6$. Our data are consistent with the predicted overhead factor of $1.56$
relative to the asymptotic Heisenberg-limited standard deviation $\pi/N$.
Despite this constant overhead, our phase estimates clearly surpass the SQL. For
example, where we have demonstrated the use of 378 resources in our algorithm
(corresponding to a maximum of 32 passes), 4,333 resources would be required
using standard techniques to achieve the same uncertainty.

We have introduced a new algorithm for phase estimation, generalizing Kitaev's
algorithm, which requires no entanglement to achieve Heisenberg-limited scaling
independent of $\phi$. Our algorithm uses single-photon Fock states, multiple
passes and adaptive measurement. We have used our algorithm to successfully
demonstrate the first measurement with Heisenberg-limited scaling. This
technique has promise for a wide range of metrology tasks, especially in light
of continued development of high-flux single photon sources and efficient
detectors.

\section*{Methods Summary}

\textbf{Interferometer.}
A spontaneous parametric downconverter supplies pairs of single photons, one to
the interferometer and one to an independent detector. The state after the
polarizing beam splitter is $(\ket R + \ket L)/\sqrt 2$, equivalent to the
logical state $H\ket 0$. A photon passing through the 50-mm-diameter
$\phi$ phase shift half-wave plate undergoes a polarization rotation:
$\tfrac{1}{\sqrt{2}}\left(\ket R + \ket L \right) \rightarrow
\tfrac{1}{\sqrt{2}}\left(\ket R + e^{i\phi} \ket L \right)$ for a half-wave
plate setting of $\phi/4$. Two 50-mm-diameter mirrors are placed on
either side, allowing a single photon to pass through the half-wave plate
multiple times. To correct the unwanted $\pi$-phase shift (in the $\ket H /
\ket V$ basis) on reflection, a quarter-wave plate, set to its optic axis, is
inserted before each of the large mirrors. The feedback phase is implemented as
another half-wave plate mounted in a computer-controlled rotation stage before
the unknown phase and mirrors. We use a fixed phase $\phi$, but the use of a
uniformly distributed random initial feedback phase is equivalent to performing
the protocol over the full range of system phases, $\phi\in\left[0,2\pi\right)$.
Mirrors on computer-controlled translation stages are used to select the $2^k$th
pass for each value of $k$. Measurement is performed in the horizontal/vertical
basis, corresponding to a $\op\sigma_x$ measurement, with a high-contrast-ratio
calcite polarizing beam displacer. The two outputs of the beam displacer,
filtered with 10-nm-bandwidth filters to reject background light, are
sent to single photon counting modules. A successful measurement is
heralded by a coincidence between the directly detected photon and either output
detector.

\textbf{Statistics for phase.}
For a phase $\phi$ with distribution $P$, an appropriate measure of error in the
estimate $\phi_\text{est}$ is $ \langle
\cos\left(\phi-{\phi}_\text{est}\right)\rangle_P^{-2}-1$. This achieves its
minimum, the Holevo variance $V_\text{H}$ \cite{Wiseman1997},
for ${\phi}_\text{est} = \arg \left( \langle \exp\left(i\phi\right)
\rangle_P\right)$, which is the estimate we use. When applied to a phase the
terms variance and standard deviation are to be understood as
$V_\text{H}$ and $\sqrt{V_\text{H}}$ respectively.

\section*{Methods}

\textbf{Source.}
Our type-I BiBO (bismuth borate) spontaneous parametric down-conversion source
is pumped by a frequency-doubled mode-locked Ti:sapphire laser, producing pairs
of 820-nm, 2-nm-bandwidth single photons in the state $\ket{HH}$. One photon is
guided to the experiment through a single-mode optical fibre, the other is
guided straight to a single-photon counting module. A successful measurement is
heralded by a coincidence between the directly detected photon and either of the
output detectors---coincidence detection reduces background and dark counts,
ensuring high-fidelity conditional single photons in the experiment.

\textbf{Quarter-wave plate setting.}
For logistical reasons, we use 25-mm-diameter quarter-wave plates for most
experiments, but 50-mm-diameter wave plates for the $K=5$ cases. The
wave plates are nominally identical except for their diameter. The quarter-wave
plates did introduce a technical challenge: because of the large-diameter
optics, and the need to use all of the clear aperture to obtain multiple
reflections, we used mountings that did not allow easy calibration and
adjustment of the quarter-wave plates. This in turn led to small additional
phase shifts from the quarter-wave plates that were dependent upon the number of
passes. This problem is easily modelled and is not fundamental.

\textbf{Analytic solutions.}
For the cases $M=1$ (Kitaev) and $M=2$ of our algorithm we have shown
analytically that the variance scales as the SQL, by solving the adaptive scheme
exactly, using the formulae in Ref. \cite{Berry2000}. The exact results for the
variances are $2/N+1/N^2$ and $2/N$ respectively.

\textbf{Error calculation.}
The 95\% confidence intervals shown were determined using a studentized
bootstrap on a log scale. The bootstrap is a method of determining confidence
intervals without making assumptions about the form of the underlying
distribution \cite{Davison1997}. The data are used as a model of the underlying
distribution, and confidence intervals for the quantity of interest are
estimated by sampling from this distribution. That is, a number of subsamples
equal to the size of the data set, $m$, is obtained, and an estimate of the
quantity of interest is determined from this set of subsamples. A large
number of bootstrap samples are used (in our case $10^6-1$), where each sample
is an estimate of the quantity of interest based on the set of $m$ subsamples.
The distribution obtained for the quantity of interest is then used to determine
a confidence interval, as described on page 199 of Ref.~\cite{Davison1997}.

For accuracy, the quantity of interest should have an uncertainty which is
independent of the value of the quantity. As the uncertainty in an estimate of
variance is approximately proportional to the variance, taking the logarithm of
the variance yields a quantity with constant uncertainty (as is done, for
example, in Ref. \cite{Schenker1985}). The studentized bootstrap yields
additional accuracy, and involves normalizing by the estimated uncertainty in
the quantity of interest \cite{Davison1997}. We have used both used the log
scale and the studentized bootstrap in order to obtain accurate error bars.

\textbf{Acknowledgements} We thank M. Mitchell, D. Bulger and S. Lo for discussions. This work was supported by the Australian Research Council and the Queensland State Government.

\textbf{Author information} Correspondence should be addressed to G.J.P. (email: g.pryde@griffith.edu.au).

\end{document}